\documentclass[reprint, amsmath,amssymb, aps, superscriptaddress, floatfix]{revtex4-1}

\usepackage[utf8]{inputenc}
\usepackage{graphicx}
\usepackage{dcolumn}
\usepackage{bm}
\usepackage{braket}
\usepackage{color}
\usepackage{units}
\usepackage{dsfont}
\usepackage{float}
\usepackage{hyperref}
\usepackage{xcolor}
\usepackage{comment}

\begin{document}

\title{Programmable Photonic Quantum Circuits with Ultrafast Time-bin Encoding}

\author{Fr\'ed\'eric Bouchard}
\email{frederic.bouchard@nrc-cnrc.gc.ca}
\affiliation{National Research Council of Canada, 100 Sussex Drive, Ottawa, Ontario K1A 0R6, Canada}
\author{Kate Fenwick}
\affiliation{National Research Council of Canada, 100 Sussex Drive, Ottawa, Ontario K1A 0R6, Canada}
\affiliation{Department of Physics, University of Ottawa, Advanced Research Complex, 25 Templeton Street, Ottawa ON Canada, K1N 6N5}
\author{Kent Bonsma-Fisher}
\affiliation{National Research Council of Canada, 100 Sussex Drive, Ottawa, Ontario K1A 0R6, Canada}
\author{Duncan England}
\affiliation{National Research Council of Canada, 100 Sussex Drive, Ottawa, Ontario K1A 0R6, Canada}
\author{Philip J. Bustard}
\affiliation{National Research Council of Canada, 100 Sussex Drive, Ottawa, Ontario K1A 0R6, Canada}
\author{Khabat Heshami}
\affiliation{National Research Council of Canada, 100 Sussex Drive, Ottawa, Ontario K1A 0R6, Canada}
\affiliation{Department of Physics, University of Ottawa, Advanced Research Complex, 25 Templeton Street, Ottawa ON Canada, K1N 6N5}
\author{Benjamin Sussman}
\affiliation{National Research Council of Canada, 100 Sussex Drive, Ottawa, Ontario K1A 0R6, Canada}
\affiliation{Department of Physics, University of Ottawa, Advanced Research Complex, 25 Templeton Street, Ottawa ON Canada, K1N 6N5}

\begin{abstract}
We propose a quantum information processing platform that utilizes the ultrafast time-bin encoding of photons. This approach offers a pathway to scalability by leveraging the inherent phase stability of collinear temporal interferometric networks at the femtosecond-to-picosecond timescale. The proposed architecture encodes information in ultrafast temporal bins processed using optically induced nonlinearities and birefringent materials while keeping photons in a single spatial mode. We demonstrate the potential for scalable photonic quantum information processing through two independent experiments that showcase the platform's programmability and scalability, respectively. The scheme's programmability is demonstrated in the first experiment, where we successfully program 362 different unitary transformations in up to 8 dimensions in a temporal circuit. In the second experiment, we show the scalability of ultrafast time-bin encoding by building a passive optical network, with increasing circuit depth, of up to 36 optical modes. In each experiment, fidelities exceed 97\%, while the interferometric phase remains passively stable for several days.
\end{abstract}

\maketitle

Coherence lies at the heart of the nonclassical quantum phenomena - entanglement and superposition - that underpin quantum technologies in computing, sensing, and communications. Decoherence, in which information leaks from a target system to its environment, is a persistent obstacle to the practical realization of these transformative capabilities. Realizing states that exhibit sustained quantum coherence necessitates a careful selection of both the physical system and encoding scheme. In comparison to many systems, photons exhibit a unique resistance to decoherence, maintaining their quantum state over extended periods of time and distances, even in ambient conditions. This robustness, coupled with their relative ease of control and detection, renders them exceptional candidates for foundational investigations of quantum mechanics~\cite{shalm2015strong,giustina2015significant} and quantum technology applications~\cite{knill2001scheme}. A particularly promising strategy lies in the development of photonic quantum systems that operate on ultrafast timescales, as short as femtoseconds. Such a rapid operational timeline ensures that quantum information processing tasks are completed before decoherence can significantly disrupt the system, thereby preserving the integrity of the quantum information. However, maintaining coherence between different photonic states is non-trivial, owing to the challenges of maintaining phase stability between distinct paths. Here we present a platform that makes use of ultrafast time-bin encoded photons, propagating on a collinear path, to perform quantum processing.

\begin{figure*}[t!]
	\centering
		\includegraphics[width=0.95\textwidth]{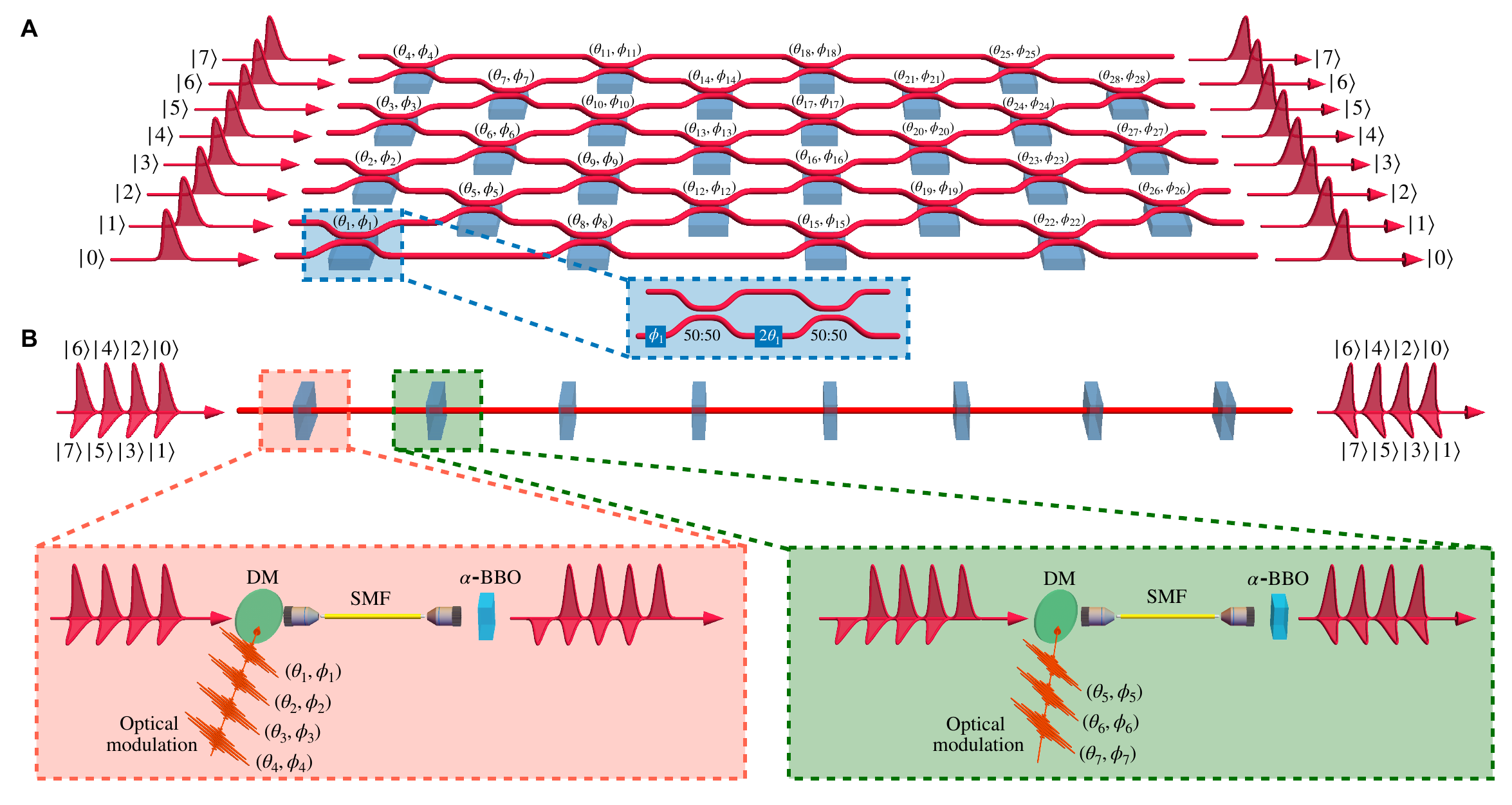}
	\caption{\textbf{Quantum circuits comparison}. \textbf{A} Typically, an array of photonic quantum states is processed by a linear optical network in a waveguide configuration. The unitary transformation is programmed by tuning a sequence of beam splitters and phase shifters or by tuning two independent phase shifters in a balanced Mach-Zehnder interferometer. The output of the quantum circuit is then directed to the detection stage. \textbf{B} Here, an ultrafast pulse train of photonic quantum states is processed by a temporal linear optical network in a single spatial mode. The optical Kerr effect modulation, accomplished by overlapping bright control pulses with the quantum signals inside a single-mode fiber, applies phase shifts and couples modes. Birefringent $\alpha$-BBO crystals act as time-bin interferometers. The output of the ultrafast temporal quantum circuit is then directed to the detection stage. DM: Dichroic Mirror; SMF: single-mode fiber.}
	\label{fig:QCircuit}
\end{figure*}
 
Generating and manipulating quantum states of light, from single photons to complex multi-photon states, remains a significant challenge. To harness the full potential of photonic platforms, photonic quantum processors must be capable of generating and controlling these states across a large number of optical modes. Recent advances in integrated programmable circuits have enabled universal programmability in up to 20 modes~\cite{carolan2015universal,harris2017quantum,taballione20198,taballione2021universal,arrazola2021quantum,taballione202220}. Although significant progress has been made in the development of fully-programmable universal photonic quantum processors, scaling these technologies to accommodate and manipulate a larger number of modes remains an open challenge. Spatial encoding in integrated and bulk photonic platforms currently serve as the backbone of most quantum photonic architectures; however, another approach using time-bin encoding, or time-multiplexing, is emerging with the potential for low loss and programmability~\cite{yokoyama2013ultra,humphreys2013linear,motes2014scalable,he2017time,asavanant2019generation,madsen2022quantum,sempere2022experimentally}.
One of the critical requirements in scaling quantum photonic platforms will be maintaining interferometric phase stability across the entirety of the quantum device and over extended periods of time. This challenge is particularly important when dealing with multi-photon states, given the heightened phase sensitivity of highly-entangled states with large photon numbers. This is famously demonstrated by N00N states~\cite{afek2010high} and other highly-entangled multi-photon states~\cite{guo2020distributed}. In addition to phase stability, other key requirements for developing photonic quantum processors include high modulation speeds, high efficiencies, high fidelities, low noise levels, and seamless integration with the source and detectors. Our approach, as we will demonstrate, offers a promising path to meeting all of these requirements.

This work proposes and demonstrates an experimental platform to universal photonic quantum information processing through ultrafast time-bin encoding (UTBE). The platform’s building blocks are a series of programmable optical ultrafast temporal interferometric network elements (POUTINEs) that can be cascaded to form a fully connected temporal quantum circuit. Time-bin photonic states are prepared, and evolve, in a single beam.  We program phase shifting and mode coupling operations by a combination of passive birefringent elements and active modulation through nonlinear optical Kerr effect interactions with intense ultrafast control pulses in optical fiber~\cite{england2021perspectives}. The fiber-based architecture of our platform offers compatibility with a wide range of quantum sources and new detector technologies, such as superconducting nanowire detectors and photon-number resolving detectors.

Perhaps the key attribute of our approach is the inherent phase stability of temporal modes at this timescale~\cite{donohue2013coherent,bouchard2022quantum,bouchard2023measuring}, a significant benefit when seeking to upscale processing to larger circuits. The ultrafast time-bin interferometers are extremely compact, robust and can operate in a single spatial mode. The birefringent elements required to create interferometric pathways between distinct time bins provide excellent interferometric phase stability. This raises the possibility of upscaling critical quantum technologies such as quantum computing, quantum simulation, quantum communication, and quantum sensing. In what follows, we showcase the versatility and potential of our platform through two separate experiments that highlight its programmability, scalability, and stability.

\begin{figure*}[t!]
	\centering
		\includegraphics[width=0.95\textwidth]{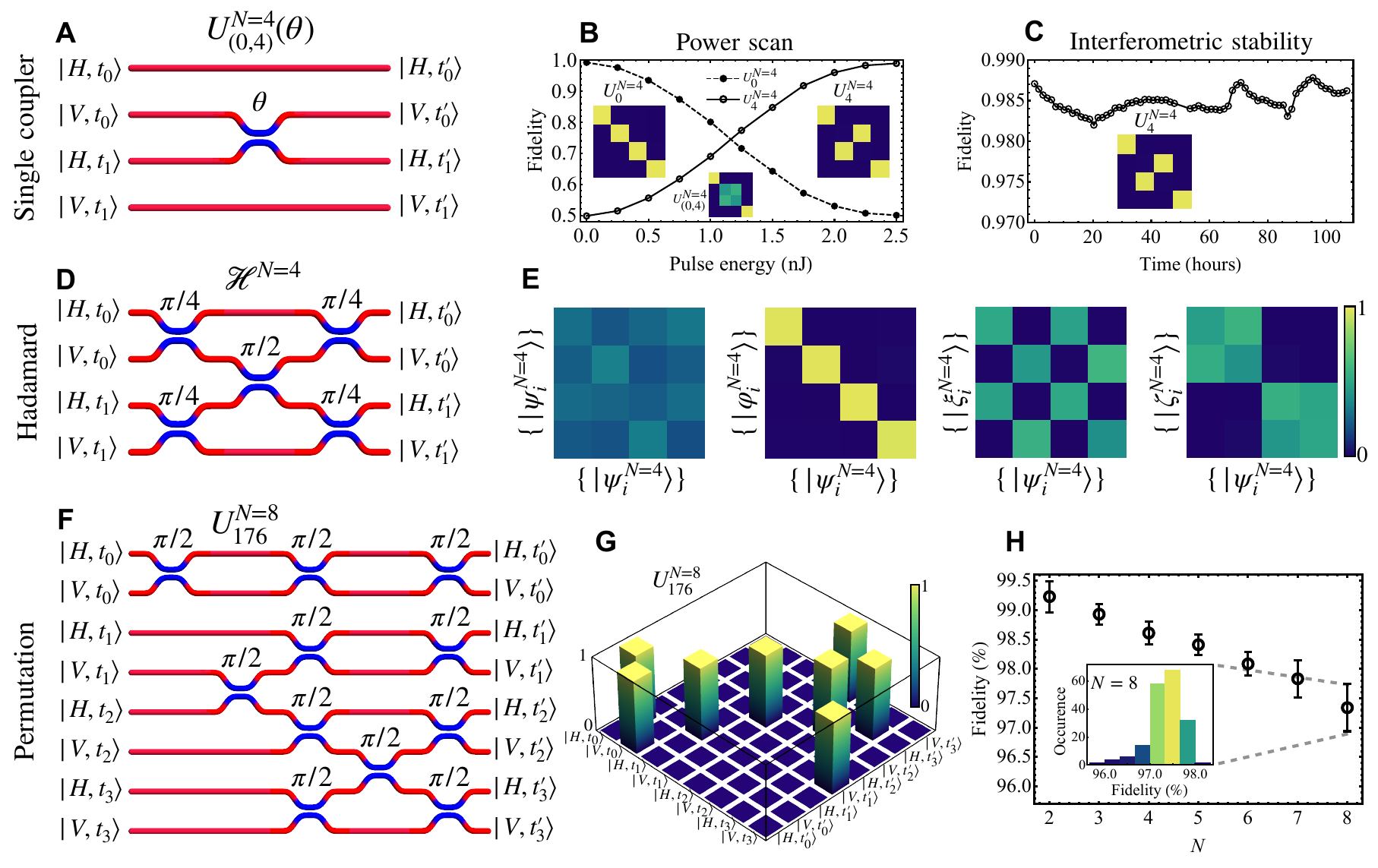}
	\caption{\textbf{Experimental demonstration of the programmable quantum temporal circuit}. 
 This figure presents a range of experimental results that showcase the programmability, robustness, and scalability of our UTBE platform for linear optical quantum computing. \textbf{A} shows the circuit diagram of a 4-dimensional temporal circuit with a mode coupler, where the coupling parameter $\theta$ controls the transformation from the identity $U_0^{N=4}$ to a SWAP-like gate $U_4^{N=4}$. {This transformation is achieved with an overall system loss, including detector efficiency, of -5.2~dB.} \textbf{B} depicts the experimental measurement of the temporal circuit's evolution and fidelity with respect to $\theta$, while \textbf{C} shows the long-term robustness of the SWAP-like gate without active stabilization or any other intervention. \textbf{D} illustrates the 4-dimensional Hadamard gate with the corresponding $\theta$ values, while \textbf{E} shows the measured probability of detection matrices of the Hadamard gate for different input bases. {This transformation is achieved with an overall system loss, including detector efficiency, of -7.2~dB.} \textbf{F} displays an example of an $N=8$ permutation transformation ($U_{176}^{N=8}$) with each mode coupler's coupling parameter. The experimentally measured detection probability matrix is shown in \textbf{G}. {This transformation is achieved with an overall system loss, including detector efficiency, of -7.2~dB.} \textbf{H} demonstrates the mean fidelity of all 362 measured permutation transformations from $N=2$ to $N=8$. For $N=8$, the fidelity distribution is shown explicitly in the inset. 
 }
	\label{fig:GatesResults}
\end{figure*}

Our goal is to build a universal linear optical network, described by an $N \times N$ unitary matrix, capable of transforming a set of $N$ input time-bin modes into any arbitrary set of $N$ output modes~\cite{reck1994experimental,clements2016optimal}. With integrated waveguides~\cite{spring2013boson,anton2019interfacing,taballione2021universal} or bulk optical devices~\cite{wang2019boson,zhong2021phase}, such a linear optical network consists of a large multiport interferometer with $O(N^2)$ points of interference that must be carefully tuned and stabilized, see Fig.~\ref{fig:QCircuit}-\textbf{A}. {To retain the full programmability of our circuits, $O(N^2)$ control pulses are required. Nevertheless, the proposed ultrafast time-bin architecture requires only $O(N)$ components, i.e. POUTINEs, cascaded in a single beam line, see Fig.~\ref{fig:QCircuit}-\textbf{B}.} The $N$-dimensional encoding space is defined using $N$ creation operators, $\{ \hat{a}_{H,t_0}^\dagger, \hat{a}_{V,t_0}^\dagger, \hat{a}_{H,t_1}^\dagger, \hat{a}_{V,t_1}^\dagger, ..., \hat{a}_{H,t_{N/2}}^\dagger, \hat{a}_{V,t_{N/2}}^\dagger \}$, associated with horizontal ($H$) and vertical ($V$) polarization modes and ultrafast time bins, $t_i$, where ${i\in {0,...,N/2}}$. In the computational basis, single-photon states are represented by $\{|\psi_i^N \rangle \} = \{(1\ 0)^T \otimes |t_0 \rangle, (0\ 1)^T \otimes |t_0 \rangle, ... , (1\ 0)^T \otimes |t_{N/2} \rangle, (0\ 1)^T \otimes |t_{N/2} \rangle \}$, for even $N$, where $(1\ 0)^T$ and $(0 \ 1)^T$ correspond to horizontally and vertically polarized states, respectively. The overall unitary transformation is achieved through a sequence of ${N(N-1)/2}$ programmable mode couplers, analogous to beam splitters and phase shifters, acting on adjacent modes in the network. 

 Our approach utilizes mode couplers that manipulate polarization at a single time bin, implemented via the optical Kerr effect, a third-order nonlinearity that induces local birefringence through cross-phase modulation with a bright pulse of light~\cite{kupchak2019terahertz},
\begin{eqnarray}
&&U_\mathrm{Kerr}^{(k)}(\theta_{m}, \phi_{m} )  \\ && = \sum_{j}
\begin{pmatrix} e^{i \phi_{m(j,k)}} \cos \theta_{m(j,k)} & - \sin \theta_{m(j,k)}\\
e^{i \phi_{m(j,k)}} \sin \theta_{m(j,k)} & \cos \theta_{m(j,k)}
\end{pmatrix} \otimes |t_j\rangle \langle t_j|, \nonumber
\end{eqnarray}
where $\theta_m$ and $\phi_m$ are the coupling and phase shifting parameters, respectively, $k\in \{0,...,N-1\}$, $j\in \{0,...,(N+(-1)^k-1)/2\}$, and $m\in \{1,...,N(N-1)/2\}$, for even $N$. For adjacent modes within a time bin, i.e. $\hat{a}_{H,t_j}^\dagger$ and $\hat{a}_{V,t_j}^\dagger$, this sequence of induced polarization rotations, $U_\mathrm{Kerr}(\theta_m,\phi_m)$, directly achieves the required mode coupling. However, for adjacent modes belonging to different time bins, i.e. $a_{V,t_j}^\dagger$ and $a_{H,t_{j+1}}^\dagger$, a birefringent crystal is required to bring both modes back to a single time bin, i.e.,
\begin{eqnarray}
U_\mathrm{Bir} = \sum_{j=0}^{N/2} \left( 
\begin{pmatrix} 1 & 0\\
0 & 0
\end{pmatrix} \otimes |t_j\rangle \langle t_j| + 
\begin{pmatrix} 0 & 0\\
0 & 1
\end{pmatrix} \otimes |t_{j+1}\rangle \langle t_j| \right)
\end{eqnarray}
{The birefringent crystal used in our scheme acts as a time-bin interferometer, similar to unbalanced Mach-Zehnder or fibre loop interferometers, where the path difference between both arms of the interferometer matches the time-bin separation. This is achieved by using birefringent crystals of the same length as those used in the generation of the ultrafast time-bin states.} In order to bring adjacent modes within a time bin back together, we employ another transformation, namely a polarization rotation given by,
\begin{eqnarray}
R(\Theta) = 
\begin{pmatrix} \cos \Theta & - \sin \Theta\\
\sin \Theta & \cos \Theta 
\end{pmatrix} \otimes \sum_{j=0}^{N/2} |t_j\rangle \langle t_j|,
\end{eqnarray}
which can be achieved with waveplates or by rotating the subsequent birefringent crystal. A single POUTINE is thus given by
\begin{eqnarray}
 {\cal U}^{(k)}(\theta_m,\phi_m) = R(\pi/2) \cdot U_\mathrm{Bir} \cdot U_\mathrm{Kerr}^{(k)} (\theta_m,\phi_m).
 \end{eqnarray}
Finally, the overall unitary transformation can then be constructed from a sequence of $N$ POUTINEs, i.e. $U^{(N)}(\theta_m,\phi_m)=\Pi_{k=0}^{N-1} {\cal U}^{(k)} (\theta_m,\phi_m)$.

The use of picosecond time-bins is a crucial advantage in our approach as it enables the implementation of highly compact time-bin interferometers with sub-millimeter path difference. This can be achieved by collinearly propagating two orthogonal linear polarization modes through a few millimeters of birefringent crystal. The compact size of these interferometers represents a significant improvement over alternative methods and reduces the experimental overhead. For instance, in previous demonstrations, time-bin interferometers have been realized with path differences reaching several kilometers, making the phase locking of such interferometers a significant achievement~\cite{madsen2022quantum}. However, the use of ultrafast time bins in our platform eliminates this requirement altogether. 

\begin{figure*}[t!]
	\centering
		\includegraphics[width=0.95\textwidth]{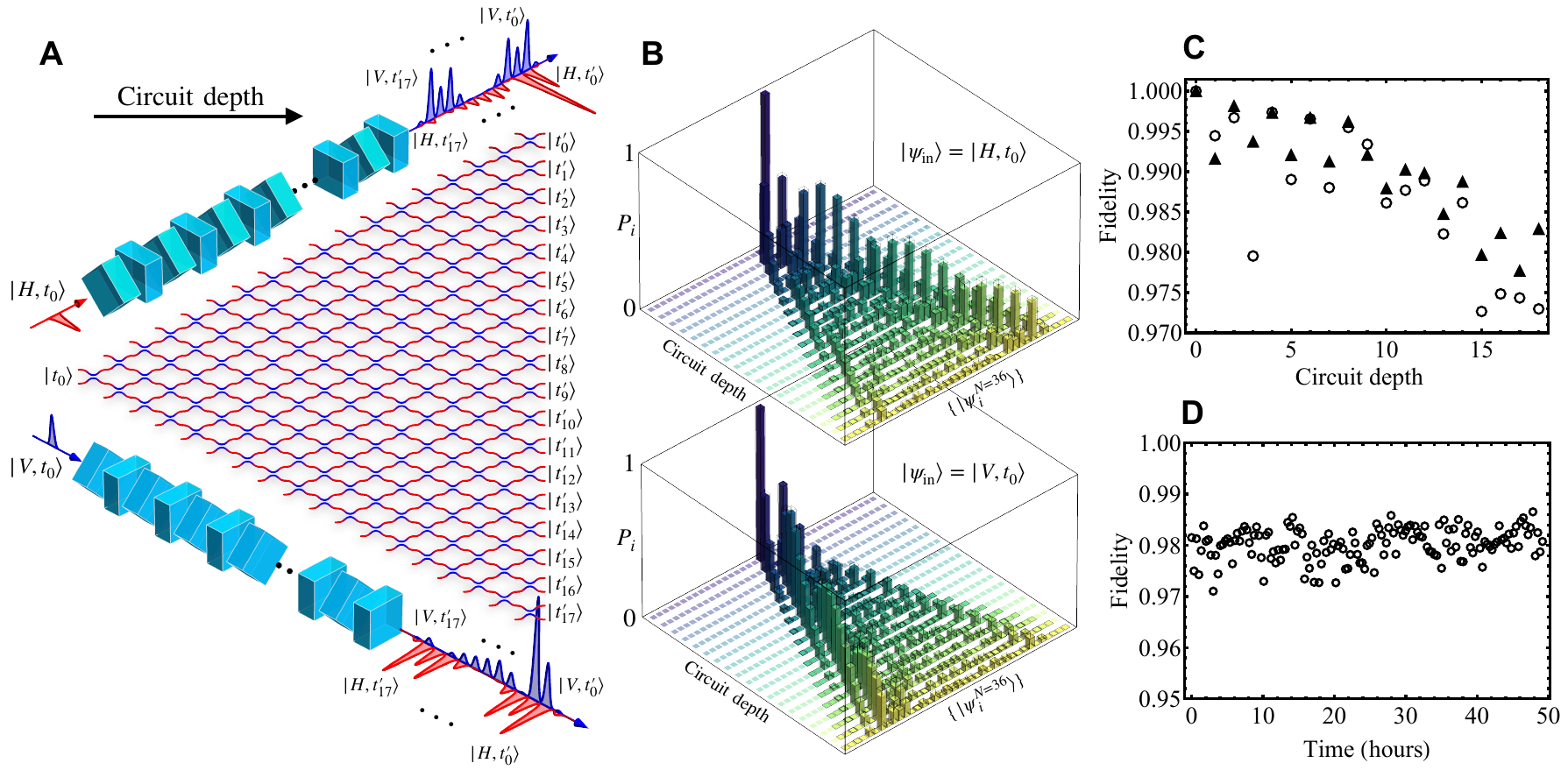}
	\caption{\textbf{Scalability demonstration of the programmable quantum temporal circuit}. 
 \textbf{A} shows the theoretical output probability distribution of a fixed temporal circuit with 171 couplers and varied circuit depth for different input states, $|\psi_\mathrm{in} \rangle = |H,t_0\rangle$ (top) and $|\psi_\mathrm{in} \rangle = |V,t_0\rangle$ (bottom). \textbf{B} displays the measured output state {probability} compared to the target state (black outline) for both input states as a function of circuit depth. \textbf{C} shows the fidelity as a function of circuit depth for the two input states (triangles and circles corresponding to horizontal and vertical input polarization states, respectively). \textbf{D} shows the long-term stability of the $N=36$ temporal circuit with a circuit depth of 18, with no significant decrease in output state fidelity observed over a period of 50 hours. {This transformation is achieved with an overall system loss, including detector efficiency, of -13~dB.}
 }
	\label{fig:WalkResults}
\end{figure*}

To assess the performance of our platform, we began by designing a simple temporal circuit with $N=4$ optical modes comprised of a single optically induced mode coupler, as depicted in Fig.\ref{fig:GatesResults}-\textbf{A}. Using a Ti:Sa laser to pump an optical parametric oscillator at a rate of 80~MHz, we generate single-photon-level signal pulses and introduce them into the temporal circuit. {Here, we use $\alpha$-BBO crystals with a thickness of 10~mm as our birefringent elements to create the appropriate ultrafast time bins with a corresponding temporal separation of 4.3~ps. Details about the generation and detection of ultrafast time-bin states can be found in the Supplemental Material.} Control pulses are then prepared from the pump laser to program the circuit, while taking care to select non-overlapping wavelengths for the signal and control pulses, i.e. $\lambda_\mathrm{signal}=715~\mathrm{nm}$ and $\lambda_\mathrm{control}=790~\mathrm{nm}$. { The average efficiency of a single POUTINE is given by -1~dB of loss and is currently limited by the coupling efficiency to the single-mode fiber.} The mode coupling parameter, $\theta$, is controlled by varying the energy of the control pulse, allowing us to continuously tune the circuit from the identity transformation, $U_0^{N=4}$, to a SWAP-like gate, $U_4^{N=4}$, as shown in Fig.\ref{fig:GatesResults}-\textbf{B}. To assess the quality and stability of the circuit, we measure the output number of counts distribution for each input mode and compute the fidelity, $F=(1/N)\, \mathrm{Tr}\left( | U_\mathrm{target}^\dagger | \cdot | U_\mathrm{exp} | \right)$, where $U_\mathrm{target}$ and $U_\mathrm{exp}$ are the target and experimentally measured probability of detection matrices, respectively. Although interferometric phase stability can be a challenge for linear optical networks, even for those as small as $N=4$, we found no degradation or significant drift over several days, see Fig.\ref{fig:GatesResults}-\textbf{C}. The small variations in fidelity we observed were likely due to fluctuations in the control pulse energy. Our ultra-stable interferometric phase arises from the fact that the phase of each arm of the multiport time-bin interferometer is determined by the tilting angle of the birefringent crystals, which can be kept fixed with high precision. We ended our stability measurements after 108 hours, but expect the temporal circuit to remain phase stable for much longer. 

Our exploration of temporal quantum circuits next progresses to a more complex network, the high-dimensional Hadamard gate. Although experimentally challenging to implement, high-dimensional Hadamard gates are an important component of qudit-based quantum computing approaches~\cite{brandt2020high,chi2022programmable,ringbauer2022universal}. To the best of our knowledge, this is the first experimental demonstration of a high-dimensional Hadamard gate implemented in the temporal degree of freedom of photons. The circuit diagram for the four-dimensional Hadamard gate, ${\cal H}^{N=4}$, is depicted in Fig.~\ref{fig:GatesResults}-\textbf{D}, which can be realized with two control pulses and half-wave plates acting uniformly across all time bins in the temporal network. We characterized the quality of the Hadamard circuit by measuring the probability of detection matrices for input quantum signals prepared in different bases, including the computational basis, $\{|\psi_i^{N=4} \rangle \}=\{|H,t_0\rangle,|V,t_0\rangle,|H,t_1\rangle,|V,t_1\rangle \}$, the four-dimensional discrete Fourier transform basis, $\{|\varphi_i^{N=4} \rangle \}=\{ (|H\rangle \pm |V\rangle) \otimes (|t_0\rangle \pm |t_1\rangle )/2 \}$, and other superposition bases, such as $\{ |\xi_i^{N=4} \rangle \} = \{ (|H\rangle \pm |V\rangle) \otimes |t_0\rangle / \sqrt{2} , (|H\rangle \pm |V\rangle) \otimes |t_1\rangle /\sqrt{2} \} $ and $\{ | \zeta_i^{N=4} \rangle \} = \{ |H\rangle \otimes (|t_0\rangle \pm |t_1\rangle) / \sqrt{2}, |V\rangle \otimes (|t_0\rangle \pm |t_1\rangle)/\sqrt{2} \} $.

To further demonstrate the versatility and controllability of our temporal quantum processor, we perform a set of high-dimensional permutation transformations, which are essential for quantum information processing. The number of distinct permutation transformations for a linear optical network with $N$ modes is $N!$. In our experiments, we were able to implement and measure all 32 permutation matrices for $N$ up to $4$, while for larger dimensions, we sampled a subset of permutations due to their overwhelming number. In total, we have experimentally implemented and measured 362 permutation transformations for dimensions ranging from $N=2$ to $N=8$. As an illustration, Fig.~\ref{fig:GatesResults}-\textbf{F} and \textbf{G} respectively show the circuit diagram and the probability of detection matrix obtained experimentally for the $N=8$ permutation transformation $U_{176}^{N=8}$. We achieved an average transformation fidelity above 97\% for all measured permutation matrices for all values of $N$ (Fig.~\ref{fig:GatesResults}-\textbf{H}). The small decrease in fidelity can be attributed to an increase in total noise when considering a larger number of modes.

Having demonstrated the programmability of our platform by achieving high-dimensional unitary transformations in the temporal degree of freedom, we next investigate phase stability when scaling temporal circuits to a large number of modes, as illustrated in Fig.\ref{fig:WalkResults}-\textbf{A}. For this purpose, we consider a fixed circuit with each coupler set to $\theta=\pi/4$ across all time bins, and measure the output distribution of the circuit when fed with a single time-bin for increasing circuit depths, as shown in Fig.\ref{fig:WalkResults}-\textbf{B}. Remarkably, even for the largest circuit depth ($N=36$) with a total of 171 mode couplers involved in the transformation, the same prolonged phase stability observed in the previous demonstrations ($N=2,..,8$) is still maintained over an extended period of time, i.e., over 50 hours for $N=36$, as seen in Fig.\ref{fig:WalkResults}-\textbf{C}. Moreover, the transformation fidelity remains excellent even as we increase the circuit depth, as depicted in Fig.\ref{fig:WalkResults}-\textbf{D}. Notably, since the 36-mode temporal interferometric setup occurs collinearly in a single spatial mode, the number of physical elements scales linearly with circuit depth, and only 18 birefringent crystals are required to construct this large-scale circuit.

Our results demonstrate the significant potential of the UTBE platform for advancing quantum technologies. Although our platform is based on polarization to create the temporal circuit, other degrees of freedom such as frequency-bins or path-encoding can be used in parallel with the ultrafast time-bins. Frequency conversion of quantum signals based on third-order nonlinearities, such as cross-phase modulation~\cite{matsuda2016deterministic} and four-wave mixing~\cite{mcguinness2010quantum}, have recently seen remarkable progress, enabling further development of frequency-bin encoding~\cite{joshi2018frequency}. Compared to polarization, a hybrid time-bin and frequency-bin encoding can offer a much larger encoding space, extending to multiple dimensions with fewer experimental resources. In this scheme, time-bin interferometers can be realized by replacing the birefringent crystals with dispersion-engineered fibers, which provide greater control and programmability of ultrafast temporal quantum processors. Furthermore, integrating the UTBE architecture into a waveguide architecture with ultrafast control could leverage past investments in integrated photonics and exploit key features of our ultrafast platform. Finally, recent advances in superconducting nanowire single-photon detectors, where a sub-3~ps temporal resolution has been achieved~\cite{korzh2020demonstration}, enable efficient measurements of ultrafast time-bin encoded photons.

In conclusion, the proposed platform presents a promising scalability strategy that prioritizes phase stability as a critical feature to develop future implementations with larger mode and photon numbers. We have demonstrated that a programmable quantum circuit with a non-trivial number of modes can be implemented without the need for complex fabrication processes. Leveraging readily available experimental components in quantum optics labs, e.g. ultrafast lasers, optical fibers, and birefringent crystals, we have provided a versatile set of tools to design and program quantum circuits, enabling researchers to explore a wide range of quantum processing applications. We anticipate that our platform will be a fertile ground for exploring novel sources, implementations, detection schemes, and other applications, such as quantum neural networks~\cite{steinbrecher2019quantum}, quantum metrology~\cite{matthews2016towards}, and quantum networks~\cite{cacciapuoti2019quantum}. UTBE is a powerful addition to current approaches as it offers a single underlying platform for generating non-classical light and conducting quantum processing. Notably, our platform has tremendous potential for enabling various applications in multi-photon entangled state generation. Moreover, the universal nature of our UTBE scheme supports a range of photonic quantum computing approaches, from non-universal approaches like Gaussian Boson sampling to full-fledged linear optical quantum computing.

\section*{Acknowledgments}
We thank Guillaume Thekkadath, Aaron Goldberg, Yingwen Zhang, Andrew Proppe, Rune Lausten, Denis Guay, and Doug Moffatt for support and insightful discussions.

\providecommand{\noopsort}[1]{}


\begin{thebibliography}{39}%
\makeatletter
\providecommand \@ifxundefined [1]{%
 \@ifx{#1\undefined}
}%
\providecommand \@ifnum [1]{%
 \ifnum #1\expandafter \@firstoftwo
 \else \expandafter \@secondoftwo
 \fi
}%
\providecommand \@ifx [1]{%
 \ifx #1\expandafter \@firstoftwo
 \else \expandafter \@secondoftwo
 \fi
}%
\providecommand \natexlab [1]{#1}%
\providecommand \enquote  [1]{``#1''}%
\providecommand \bibnamefont  [1]{#1}%
\providecommand \bibfnamefont [1]{#1}%
\providecommand \citenamefont [1]{#1}%
\providecommand \href@noop [0]{\@secondoftwo}%
\providecommand \href [0]{\begingroup \@sanitize@url \@href}%
\providecommand \@href[1]{\@@startlink{#1}\@@href}%
\providecommand \@@href[1]{\endgroup#1\@@endlink}%
\providecommand \@sanitize@url [0]{\catcode `\\12\catcode `\$12\catcode
  `\&12\catcode `\#12\catcode `\^12\catcode `\_12\catcode `\%12\relax}%
\providecommand \@@startlink[1]{}%
\providecommand \@@endlink[0]{}%
\providecommand \url  [0]{\begingroup\@sanitize@url \@url }%
\providecommand \@url [1]{\endgroup\@href {#1}{\urlprefix }}%
\providecommand \urlprefix  [0]{URL }%
\providecommand \Eprint [0]{\href }%
\providecommand \doibase [0]{http://dx.doi.org/}%
\providecommand \selectlanguage [0]{\@gobble}%
\providecommand \bibinfo  [0]{\@secondoftwo}%
\providecommand \bibfield  [0]{\@secondoftwo}%
\providecommand \translation [1]{[#1]}%
\providecommand \BibitemOpen [0]{}%
\providecommand \bibitemStop [0]{}%
\providecommand \bibitemNoStop [0]{.\EOS\space}%
\providecommand \EOS [0]{\spacefactor3000\relax}%
\providecommand \BibitemShut  [1]{\csname bibitem#1\endcsname}%
\let\auto@bib@innerbib\@empty
\bibitem [{\citenamefont {Shalm}\ \emph {et~al.}(2015)\citenamefont {Shalm},
  \citenamefont {Meyer-Scott}, \citenamefont {Christensen}, \citenamefont
  {Bierhorst}, \citenamefont {Wayne}, \citenamefont {Stevens}, \citenamefont
  {Gerrits}, \citenamefont {Glancy}, \citenamefont {Hamel}, \citenamefont
  {Allman} \emph {et~al.}}]{shalm2015strong}%
  \BibitemOpen
  \bibfield  {author} {\bibinfo {author} {\bibfnamefont {L.~K.}\ \bibnamefont
  {Shalm}}, \bibinfo {author} {\bibfnamefont {E.}~\bibnamefont {Meyer-Scott}},
  \bibinfo {author} {\bibfnamefont {B.~G.}\ \bibnamefont {Christensen}},
  \bibinfo {author} {\bibfnamefont {P.}~\bibnamefont {Bierhorst}}, \bibinfo
  {author} {\bibfnamefont {M.~A.}\ \bibnamefont {Wayne}}, \bibinfo {author}
  {\bibfnamefont {M.~J.}\ \bibnamefont {Stevens}}, \bibinfo {author}
  {\bibfnamefont {T.}~\bibnamefont {Gerrits}}, \bibinfo {author} {\bibfnamefont
  {S.}~\bibnamefont {Glancy}}, \bibinfo {author} {\bibfnamefont {D.~R.}\
  \bibnamefont {Hamel}}, \bibinfo {author} {\bibfnamefont {M.~S.}\ \bibnamefont
  {Allman}},  \emph {et~al.},\ }\href@noop {} {\bibfield  {journal} {\bibinfo
  {journal} {Physical review letters}\ }\textbf {\bibinfo {volume} {115}},\
  \bibinfo {pages} {250402} (\bibinfo {year} {2015})}\BibitemShut {NoStop}%
\bibitem [{\citenamefont {Giustina}\ \emph {et~al.}(2015)\citenamefont
  {Giustina}, \citenamefont {Versteegh}, \citenamefont {Wengerowsky},
  \citenamefont {Handsteiner}, \citenamefont {Hochrainer}, \citenamefont
  {Phelan}, \citenamefont {Steinlechner}, \citenamefont {Kofler}, \citenamefont
  {Larsson}, \citenamefont {Abell{\'a}n} \emph
  {et~al.}}]{giustina2015significant}%
  \BibitemOpen
  \bibfield  {author} {\bibinfo {author} {\bibfnamefont {M.}~\bibnamefont
  {Giustina}}, \bibinfo {author} {\bibfnamefont {M.~A.}\ \bibnamefont
  {Versteegh}}, \bibinfo {author} {\bibfnamefont {S.}~\bibnamefont
  {Wengerowsky}}, \bibinfo {author} {\bibfnamefont {J.}~\bibnamefont
  {Handsteiner}}, \bibinfo {author} {\bibfnamefont {A.}~\bibnamefont
  {Hochrainer}}, \bibinfo {author} {\bibfnamefont {K.}~\bibnamefont {Phelan}},
  \bibinfo {author} {\bibfnamefont {F.}~\bibnamefont {Steinlechner}}, \bibinfo
  {author} {\bibfnamefont {J.}~\bibnamefont {Kofler}}, \bibinfo {author}
  {\bibfnamefont {J.-{\AA}.}\ \bibnamefont {Larsson}}, \bibinfo {author}
  {\bibfnamefont {C.}~\bibnamefont {Abell{\'a}n}},  \emph {et~al.},\
  }\href@noop {} {\bibfield  {journal} {\bibinfo  {journal} {Physical review
  letters}\ }\textbf {\bibinfo {volume} {115}},\ \bibinfo {pages} {250401}
  (\bibinfo {year} {2015})}\BibitemShut {NoStop}%
\bibitem [{\citenamefont {Knill}\ \emph {et~al.}(2001)\citenamefont {Knill},
  \citenamefont {Laflamme},\ and\ \citenamefont {Milburn}}]{knill2001scheme}%
  \BibitemOpen
  \bibfield  {author} {\bibinfo {author} {\bibfnamefont {E.}~\bibnamefont
  {Knill}}, \bibinfo {author} {\bibfnamefont {R.}~\bibnamefont {Laflamme}}, \
  and\ \bibinfo {author} {\bibfnamefont {G.~J.}\ \bibnamefont {Milburn}},\
  }\href@noop {} {\bibfield  {journal} {\bibinfo  {journal} {nature}\ }\textbf
  {\bibinfo {volume} {409}},\ \bibinfo {pages} {46} (\bibinfo {year}
  {2001})}\BibitemShut {NoStop}%
\bibitem [{\citenamefont {Carolan}\ \emph {et~al.}(2015)\citenamefont
  {Carolan}, \citenamefont {Harrold}, \citenamefont {Sparrow}, \citenamefont
  {Mart{\'\i}n-L{\'o}pez}, \citenamefont {Russell}, \citenamefont
  {Silverstone}, \citenamefont {Shadbolt}, \citenamefont {Matsuda},
  \citenamefont {Oguma}, \citenamefont {Itoh} \emph
  {et~al.}}]{carolan2015universal}%
  \BibitemOpen
  \bibfield  {author} {\bibinfo {author} {\bibfnamefont {J.}~\bibnamefont
  {Carolan}}, \bibinfo {author} {\bibfnamefont {C.}~\bibnamefont {Harrold}},
  \bibinfo {author} {\bibfnamefont {C.}~\bibnamefont {Sparrow}}, \bibinfo
  {author} {\bibfnamefont {E.}~\bibnamefont {Mart{\'\i}n-L{\'o}pez}}, \bibinfo
  {author} {\bibfnamefont {N.~J.}\ \bibnamefont {Russell}}, \bibinfo {author}
  {\bibfnamefont {J.~W.}\ \bibnamefont {Silverstone}}, \bibinfo {author}
  {\bibfnamefont {P.~J.}\ \bibnamefont {Shadbolt}}, \bibinfo {author}
  {\bibfnamefont {N.}~\bibnamefont {Matsuda}}, \bibinfo {author} {\bibfnamefont
  {M.}~\bibnamefont {Oguma}}, \bibinfo {author} {\bibfnamefont
  {M.}~\bibnamefont {Itoh}},  \emph {et~al.},\ }\href@noop {} {\bibfield
  {journal} {\bibinfo  {journal} {Science}\ }\textbf {\bibinfo {volume}
  {349}},\ \bibinfo {pages} {711} (\bibinfo {year} {2015})}\BibitemShut
  {NoStop}%
\bibitem [{\citenamefont {Harris}\ \emph {et~al.}(2017)\citenamefont {Harris},
  \citenamefont {Steinbrecher}, \citenamefont {Prabhu}, \citenamefont {Lahini},
  \citenamefont {Mower}, \citenamefont {Bunandar}, \citenamefont {Chen},
  \citenamefont {Wong}, \citenamefont {Baehr-Jones}, \citenamefont {Hochberg}
  \emph {et~al.}}]{harris2017quantum}%
  \BibitemOpen
  \bibfield  {author} {\bibinfo {author} {\bibfnamefont {N.~C.}\ \bibnamefont
  {Harris}}, \bibinfo {author} {\bibfnamefont {G.~R.}\ \bibnamefont
  {Steinbrecher}}, \bibinfo {author} {\bibfnamefont {M.}~\bibnamefont
  {Prabhu}}, \bibinfo {author} {\bibfnamefont {Y.}~\bibnamefont {Lahini}},
  \bibinfo {author} {\bibfnamefont {J.}~\bibnamefont {Mower}}, \bibinfo
  {author} {\bibfnamefont {D.}~\bibnamefont {Bunandar}}, \bibinfo {author}
  {\bibfnamefont {C.}~\bibnamefont {Chen}}, \bibinfo {author} {\bibfnamefont
  {F.~N.}\ \bibnamefont {Wong}}, \bibinfo {author} {\bibfnamefont
  {T.}~\bibnamefont {Baehr-Jones}}, \bibinfo {author} {\bibfnamefont
  {M.}~\bibnamefont {Hochberg}},  \emph {et~al.},\ }\href@noop {} {\bibfield
  {journal} {\bibinfo  {journal} {Nature Photonics}\ }\textbf {\bibinfo
  {volume} {11}},\ \bibinfo {pages} {447} (\bibinfo {year} {2017})}\BibitemShut
  {NoStop}%
\bibitem [{\citenamefont {Taballione}\ \emph {et~al.}(2019)\citenamefont
  {Taballione}, \citenamefont {Wolterink}, \citenamefont {Lugani},
  \citenamefont {Eckstein}, \citenamefont {Bell}, \citenamefont {Grootjans},
  \citenamefont {Visscher}, \citenamefont {Geskus}, \citenamefont {Roeloffzen},
  \citenamefont {Renema} \emph {et~al.}}]{taballione20198}%
  \BibitemOpen
  \bibfield  {author} {\bibinfo {author} {\bibfnamefont {C.}~\bibnamefont
  {Taballione}}, \bibinfo {author} {\bibfnamefont {T.~A.}\ \bibnamefont
  {Wolterink}}, \bibinfo {author} {\bibfnamefont {J.}~\bibnamefont {Lugani}},
  \bibinfo {author} {\bibfnamefont {A.}~\bibnamefont {Eckstein}}, \bibinfo
  {author} {\bibfnamefont {B.~A.}\ \bibnamefont {Bell}}, \bibinfo {author}
  {\bibfnamefont {R.}~\bibnamefont {Grootjans}}, \bibinfo {author}
  {\bibfnamefont {I.}~\bibnamefont {Visscher}}, \bibinfo {author}
  {\bibfnamefont {D.}~\bibnamefont {Geskus}}, \bibinfo {author} {\bibfnamefont
  {C.~G.}\ \bibnamefont {Roeloffzen}}, \bibinfo {author} {\bibfnamefont
  {J.~J.}\ \bibnamefont {Renema}},  \emph {et~al.},\ }\href@noop {} {\bibfield
  {journal} {\bibinfo  {journal} {Optics express}\ }\textbf {\bibinfo {volume}
  {27}},\ \bibinfo {pages} {26842} (\bibinfo {year} {2019})}\BibitemShut
  {NoStop}%
\bibitem [{\citenamefont {Taballione}\ \emph {et~al.}(2021)\citenamefont
  {Taballione}, \citenamefont {van~der Meer}, \citenamefont {Snijders},
  \citenamefont {Hooijschuur}, \citenamefont {Epping}, \citenamefont
  {de~Goede}, \citenamefont {Kassenberg}, \citenamefont {Venderbosch},
  \citenamefont {Toebes}, \citenamefont {van~den Vlekkert} \emph
  {et~al.}}]{taballione2021universal}%
  \BibitemOpen
  \bibfield  {author} {\bibinfo {author} {\bibfnamefont {C.}~\bibnamefont
  {Taballione}}, \bibinfo {author} {\bibfnamefont {R.}~\bibnamefont {van~der
  Meer}}, \bibinfo {author} {\bibfnamefont {H.~J.}\ \bibnamefont {Snijders}},
  \bibinfo {author} {\bibfnamefont {P.}~\bibnamefont {Hooijschuur}}, \bibinfo
  {author} {\bibfnamefont {J.~P.}\ \bibnamefont {Epping}}, \bibinfo {author}
  {\bibfnamefont {M.}~\bibnamefont {de~Goede}}, \bibinfo {author}
  {\bibfnamefont {B.}~\bibnamefont {Kassenberg}}, \bibinfo {author}
  {\bibfnamefont {P.}~\bibnamefont {Venderbosch}}, \bibinfo {author}
  {\bibfnamefont {C.}~\bibnamefont {Toebes}}, \bibinfo {author} {\bibfnamefont
  {H.}~\bibnamefont {van~den Vlekkert}},  \emph {et~al.},\ }\href@noop {}
  {\bibfield  {journal} {\bibinfo  {journal} {Materials for Quantum
  Technology}\ }\textbf {\bibinfo {volume} {1}},\ \bibinfo {pages} {035002}
  (\bibinfo {year} {2021})}\BibitemShut {NoStop}%
\bibitem [{\citenamefont {Arrazola}\ \emph {et~al.}(2021)\citenamefont
  {Arrazola}, \citenamefont {Bergholm}, \citenamefont {Br{\'a}dler},
  \citenamefont {Bromley}, \citenamefont {Collins}, \citenamefont {Dhand},
  \citenamefont {Fumagalli}, \citenamefont {Gerrits}, \citenamefont {Goussev},
  \citenamefont {Helt} \emph {et~al.}}]{arrazola2021quantum}%
  \BibitemOpen
  \bibfield  {author} {\bibinfo {author} {\bibfnamefont {J.~M.}\ \bibnamefont
  {Arrazola}}, \bibinfo {author} {\bibfnamefont {V.}~\bibnamefont {Bergholm}},
  \bibinfo {author} {\bibfnamefont {K.}~\bibnamefont {Br{\'a}dler}}, \bibinfo
  {author} {\bibfnamefont {T.~R.}\ \bibnamefont {Bromley}}, \bibinfo {author}
  {\bibfnamefont {M.~J.}\ \bibnamefont {Collins}}, \bibinfo {author}
  {\bibfnamefont {I.}~\bibnamefont {Dhand}}, \bibinfo {author} {\bibfnamefont
  {A.}~\bibnamefont {Fumagalli}}, \bibinfo {author} {\bibfnamefont
  {T.}~\bibnamefont {Gerrits}}, \bibinfo {author} {\bibfnamefont
  {A.}~\bibnamefont {Goussev}}, \bibinfo {author} {\bibfnamefont {L.~G.}\
  \bibnamefont {Helt}},  \emph {et~al.},\ }\href@noop {} {\bibfield  {journal}
  {\bibinfo  {journal} {Nature}\ }\textbf {\bibinfo {volume} {591}},\ \bibinfo
  {pages} {54} (\bibinfo {year} {2021})}\BibitemShut {NoStop}%
\bibitem [{\citenamefont {Taballione}\ \emph {et~al.}(2022)\citenamefont
  {Taballione}, \citenamefont {Anguita}, \citenamefont {de~Goede},
  \citenamefont {Venderbosch}, \citenamefont {Kassenberg}, \citenamefont
  {Snijders}, \citenamefont {Smith}, \citenamefont {Epping}, \citenamefont
  {van~der Meer}, \citenamefont {Pinkse} \emph {et~al.}}]{taballione202220}%
  \BibitemOpen
  \bibfield  {author} {\bibinfo {author} {\bibfnamefont {C.}~\bibnamefont
  {Taballione}}, \bibinfo {author} {\bibfnamefont {M.~C.}\ \bibnamefont
  {Anguita}}, \bibinfo {author} {\bibfnamefont {M.}~\bibnamefont {de~Goede}},
  \bibinfo {author} {\bibfnamefont {P.}~\bibnamefont {Venderbosch}}, \bibinfo
  {author} {\bibfnamefont {B.}~\bibnamefont {Kassenberg}}, \bibinfo {author}
  {\bibfnamefont {H.}~\bibnamefont {Snijders}}, \bibinfo {author}
  {\bibfnamefont {D.}~\bibnamefont {Smith}}, \bibinfo {author} {\bibfnamefont
  {J.~P.}\ \bibnamefont {Epping}}, \bibinfo {author} {\bibfnamefont
  {R.}~\bibnamefont {van~der Meer}}, \bibinfo {author} {\bibfnamefont {P.~W.}\
  \bibnamefont {Pinkse}},  \emph {et~al.},\ }\href@noop {} {\bibfield
  {journal} {\bibinfo  {journal} {arXiv preprint arXiv:2203.01801}\ } (\bibinfo
  {year} {2022})}\BibitemShut {NoStop}%
\bibitem [{\citenamefont {Yokoyama}\ \emph {et~al.}(2013)\citenamefont
  {Yokoyama}, \citenamefont {Ukai}, \citenamefont {Armstrong}, \citenamefont
  {Sornphiphatphong}, \citenamefont {Kaji}, \citenamefont {Suzuki},
  \citenamefont {Yoshikawa}, \citenamefont {Yonezawa}, \citenamefont
  {Menicucci},\ and\ \citenamefont {Furusawa}}]{yokoyama2013ultra}%
  \BibitemOpen
  \bibfield  {author} {\bibinfo {author} {\bibfnamefont {S.}~\bibnamefont
  {Yokoyama}}, \bibinfo {author} {\bibfnamefont {R.}~\bibnamefont {Ukai}},
  \bibinfo {author} {\bibfnamefont {S.~C.}\ \bibnamefont {Armstrong}}, \bibinfo
  {author} {\bibfnamefont {C.}~\bibnamefont {Sornphiphatphong}}, \bibinfo
  {author} {\bibfnamefont {T.}~\bibnamefont {Kaji}}, \bibinfo {author}
  {\bibfnamefont {S.}~\bibnamefont {Suzuki}}, \bibinfo {author} {\bibfnamefont
  {J.-i.}\ \bibnamefont {Yoshikawa}}, \bibinfo {author} {\bibfnamefont
  {H.}~\bibnamefont {Yonezawa}}, \bibinfo {author} {\bibfnamefont {N.~C.}\
  \bibnamefont {Menicucci}}, \ and\ \bibinfo {author} {\bibfnamefont
  {A.}~\bibnamefont {Furusawa}},\ }\href@noop {} {\bibfield  {journal}
  {\bibinfo  {journal} {Nature Photonics}\ }\textbf {\bibinfo {volume} {7}},\
  \bibinfo {pages} {982} (\bibinfo {year} {2013})}\BibitemShut {NoStop}%
\bibitem [{\citenamefont {Humphreys}\ \emph {et~al.}(2013)\citenamefont
  {Humphreys}, \citenamefont {Metcalf}, \citenamefont {Spring}, \citenamefont
  {Moore}, \citenamefont {Jin}, \citenamefont {Barbieri}, \citenamefont
  {Kolthammer},\ and\ \citenamefont {Walmsley}}]{humphreys2013linear}%
  \BibitemOpen
  \bibfield  {author} {\bibinfo {author} {\bibfnamefont {P.~C.}\ \bibnamefont
  {Humphreys}}, \bibinfo {author} {\bibfnamefont {B.~J.}\ \bibnamefont
  {Metcalf}}, \bibinfo {author} {\bibfnamefont {J.~B.}\ \bibnamefont {Spring}},
  \bibinfo {author} {\bibfnamefont {M.}~\bibnamefont {Moore}}, \bibinfo
  {author} {\bibfnamefont {X.-M.}\ \bibnamefont {Jin}}, \bibinfo {author}
  {\bibfnamefont {M.}~\bibnamefont {Barbieri}}, \bibinfo {author}
  {\bibfnamefont {W.~S.}\ \bibnamefont {Kolthammer}}, \ and\ \bibinfo {author}
  {\bibfnamefont {I.~A.}\ \bibnamefont {Walmsley}},\ }\href@noop {} {\bibfield
  {journal} {\bibinfo  {journal} {Physical review letters}\ }\textbf {\bibinfo
  {volume} {111}},\ \bibinfo {pages} {150501} (\bibinfo {year}
  {2013})}\BibitemShut {NoStop}%
\bibitem [{\citenamefont {Motes}\ \emph {et~al.}(2014)\citenamefont {Motes},
  \citenamefont {Gilchrist}, \citenamefont {Dowling},\ and\ \citenamefont
  {Rohde}}]{motes2014scalable}%
  \BibitemOpen
  \bibfield  {author} {\bibinfo {author} {\bibfnamefont {K.~R.}\ \bibnamefont
  {Motes}}, \bibinfo {author} {\bibfnamefont {A.}~\bibnamefont {Gilchrist}},
  \bibinfo {author} {\bibfnamefont {J.~P.}\ \bibnamefont {Dowling}}, \ and\
  \bibinfo {author} {\bibfnamefont {P.~P.}\ \bibnamefont {Rohde}},\ }\href@noop
  {} {\bibfield  {journal} {\bibinfo  {journal} {Physical review letters}\
  }\textbf {\bibinfo {volume} {113}},\ \bibinfo {pages} {120501} (\bibinfo
  {year} {2014})}\BibitemShut {NoStop}%
\bibitem [{\citenamefont {He}\ \emph {et~al.}(2017)\citenamefont {He},
  \citenamefont {Ding}, \citenamefont {Su}, \citenamefont {Huang},
  \citenamefont {Qin}, \citenamefont {Wang}, \citenamefont {Unsleber},
  \citenamefont {Chen}, \citenamefont {Wang}, \citenamefont {He} \emph
  {et~al.}}]{he2017time}%
  \BibitemOpen
  \bibfield  {author} {\bibinfo {author} {\bibfnamefont {Y.}~\bibnamefont
  {He}}, \bibinfo {author} {\bibfnamefont {X.}~\bibnamefont {Ding}}, \bibinfo
  {author} {\bibfnamefont {Z.-E.}\ \bibnamefont {Su}}, \bibinfo {author}
  {\bibfnamefont {H.-L.}\ \bibnamefont {Huang}}, \bibinfo {author}
  {\bibfnamefont {J.}~\bibnamefont {Qin}}, \bibinfo {author} {\bibfnamefont
  {C.}~\bibnamefont {Wang}}, \bibinfo {author} {\bibfnamefont {S.}~\bibnamefont
  {Unsleber}}, \bibinfo {author} {\bibfnamefont {C.}~\bibnamefont {Chen}},
  \bibinfo {author} {\bibfnamefont {H.}~\bibnamefont {Wang}}, \bibinfo {author}
  {\bibfnamefont {Y.-M.}\ \bibnamefont {He}},  \emph {et~al.},\ }\href@noop {}
  {\bibfield  {journal} {\bibinfo  {journal} {Physical review letters}\
  }\textbf {\bibinfo {volume} {118}},\ \bibinfo {pages} {190501} (\bibinfo
  {year} {2017})}\BibitemShut {NoStop}%
\bibitem [{\citenamefont {Asavanant}\ \emph {et~al.}(2019)\citenamefont
  {Asavanant}, \citenamefont {Shiozawa}, \citenamefont {Yokoyama},
  \citenamefont {Charoensombutamon}, \citenamefont {Emura}, \citenamefont
  {Alexander}, \citenamefont {Takeda}, \citenamefont {Yoshikawa}, \citenamefont
  {Menicucci}, \citenamefont {Yonezawa} \emph
  {et~al.}}]{asavanant2019generation}%
  \BibitemOpen
  \bibfield  {author} {\bibinfo {author} {\bibfnamefont {W.}~\bibnamefont
  {Asavanant}}, \bibinfo {author} {\bibfnamefont {Y.}~\bibnamefont {Shiozawa}},
  \bibinfo {author} {\bibfnamefont {S.}~\bibnamefont {Yokoyama}}, \bibinfo
  {author} {\bibfnamefont {B.}~\bibnamefont {Charoensombutamon}}, \bibinfo
  {author} {\bibfnamefont {H.}~\bibnamefont {Emura}}, \bibinfo {author}
  {\bibfnamefont {R.~N.}\ \bibnamefont {Alexander}}, \bibinfo {author}
  {\bibfnamefont {S.}~\bibnamefont {Takeda}}, \bibinfo {author} {\bibfnamefont
  {J.-i.}\ \bibnamefont {Yoshikawa}}, \bibinfo {author} {\bibfnamefont {N.~C.}\
  \bibnamefont {Menicucci}}, \bibinfo {author} {\bibfnamefont {H.}~\bibnamefont
  {Yonezawa}},  \emph {et~al.},\ }\href@noop {} {\bibfield  {journal} {\bibinfo
   {journal} {Science}\ }\textbf {\bibinfo {volume} {366}},\ \bibinfo {pages}
  {373} (\bibinfo {year} {2019})}\BibitemShut {NoStop}%
\bibitem [{\citenamefont {Madsen}\ \emph {et~al.}(2022)\citenamefont {Madsen},
  \citenamefont {Laudenbach}, \citenamefont {Askarani}, \citenamefont
  {Rortais}, \citenamefont {Vincent}, \citenamefont {Bulmer}, \citenamefont
  {Miatto}, \citenamefont {Neuhaus}, \citenamefont {Helt}, \citenamefont
  {Collins} \emph {et~al.}}]{madsen2022quantum}%
  \BibitemOpen
  \bibfield  {author} {\bibinfo {author} {\bibfnamefont {L.~S.}\ \bibnamefont
  {Madsen}}, \bibinfo {author} {\bibfnamefont {F.}~\bibnamefont {Laudenbach}},
  \bibinfo {author} {\bibfnamefont {M.~F.}\ \bibnamefont {Askarani}}, \bibinfo
  {author} {\bibfnamefont {F.}~\bibnamefont {Rortais}}, \bibinfo {author}
  {\bibfnamefont {T.}~\bibnamefont {Vincent}}, \bibinfo {author} {\bibfnamefont
  {J.~F.}\ \bibnamefont {Bulmer}}, \bibinfo {author} {\bibfnamefont {F.~M.}\
  \bibnamefont {Miatto}}, \bibinfo {author} {\bibfnamefont {L.}~\bibnamefont
  {Neuhaus}}, \bibinfo {author} {\bibfnamefont {L.~G.}\ \bibnamefont {Helt}},
  \bibinfo {author} {\bibfnamefont {M.~J.}\ \bibnamefont {Collins}},  \emph
  {et~al.},\ }\href@noop {} {\bibfield  {journal} {\bibinfo  {journal}
  {Nature}\ }\textbf {\bibinfo {volume} {606}},\ \bibinfo {pages} {75}
  (\bibinfo {year} {2022})}\BibitemShut {NoStop}%
\bibitem [{\citenamefont {Sempere-Llagostera}\ \emph
  {et~al.}(2022)\citenamefont {Sempere-Llagostera}, \citenamefont {Patel},
  \citenamefont {Walmsley},\ and\ \citenamefont
  {Kolthammer}}]{sempere2022experimentally}%
  \BibitemOpen
  \bibfield  {author} {\bibinfo {author} {\bibfnamefont {S.}~\bibnamefont
  {Sempere-Llagostera}}, \bibinfo {author} {\bibfnamefont {R.}~\bibnamefont
  {Patel}}, \bibinfo {author} {\bibfnamefont {I.}~\bibnamefont {Walmsley}}, \
  and\ \bibinfo {author} {\bibfnamefont {W.}~\bibnamefont {Kolthammer}},\
  }\href@noop {} {\bibfield  {journal} {\bibinfo  {journal} {Physical Review
  X}\ }\textbf {\bibinfo {volume} {12}},\ \bibinfo {pages} {031045} (\bibinfo
  {year} {2022})}\BibitemShut {NoStop}%
\bibitem [{\citenamefont {Afek}\ \emph {et~al.}(2010)\citenamefont {Afek},
  \citenamefont {Ambar},\ and\ \citenamefont {Silberberg}}]{afek2010high}%
  \BibitemOpen
  \bibfield  {author} {\bibinfo {author} {\bibfnamefont {I.}~\bibnamefont
  {Afek}}, \bibinfo {author} {\bibfnamefont {O.}~\bibnamefont {Ambar}}, \ and\
  \bibinfo {author} {\bibfnamefont {Y.}~\bibnamefont {Silberberg}},\
  }\href@noop {} {\bibfield  {journal} {\bibinfo  {journal} {Science}\ }\textbf
  {\bibinfo {volume} {328}},\ \bibinfo {pages} {879} (\bibinfo {year}
  {2010})}\BibitemShut {NoStop}%
\bibitem [{\citenamefont {Guo}\ \emph {et~al.}(2020)\citenamefont {Guo},
  \citenamefont {Breum}, \citenamefont {Borregaard}, \citenamefont {Izumi},
  \citenamefont {Larsen}, \citenamefont {Gehring}, \citenamefont {Christandl},
  \citenamefont {Neergaard-Nielsen},\ and\ \citenamefont
  {Andersen}}]{guo2020distributed}%
  \BibitemOpen
  \bibfield  {author} {\bibinfo {author} {\bibfnamefont {X.}~\bibnamefont
  {Guo}}, \bibinfo {author} {\bibfnamefont {C.~R.}\ \bibnamefont {Breum}},
  \bibinfo {author} {\bibfnamefont {J.}~\bibnamefont {Borregaard}}, \bibinfo
  {author} {\bibfnamefont {S.}~\bibnamefont {Izumi}}, \bibinfo {author}
  {\bibfnamefont {M.~V.}\ \bibnamefont {Larsen}}, \bibinfo {author}
  {\bibfnamefont {T.}~\bibnamefont {Gehring}}, \bibinfo {author} {\bibfnamefont
  {M.}~\bibnamefont {Christandl}}, \bibinfo {author} {\bibfnamefont {J.~S.}\
  \bibnamefont {Neergaard-Nielsen}}, \ and\ \bibinfo {author} {\bibfnamefont
  {U.~L.}\ \bibnamefont {Andersen}},\ }\href@noop {} {\bibfield  {journal}
  {\bibinfo  {journal} {Nature Physics}\ }\textbf {\bibinfo {volume} {16}},\
  \bibinfo {pages} {281} (\bibinfo {year} {2020})}\BibitemShut {NoStop}%
\bibitem [{\citenamefont {England}\ \emph {et~al.}(2021)\citenamefont
  {England}, \citenamefont {Bouchard}, \citenamefont {Fenwick}, \citenamefont
  {Bonsma-Fisher}, \citenamefont {Zhang}, \citenamefont {Bustard},\ and\
  \citenamefont {Sussman}}]{england2021perspectives}%
  \BibitemOpen
  \bibfield  {author} {\bibinfo {author} {\bibfnamefont {D.}~\bibnamefont
  {England}}, \bibinfo {author} {\bibfnamefont {F.}~\bibnamefont {Bouchard}},
  \bibinfo {author} {\bibfnamefont {K.}~\bibnamefont {Fenwick}}, \bibinfo
  {author} {\bibfnamefont {K.}~\bibnamefont {Bonsma-Fisher}}, \bibinfo {author}
  {\bibfnamefont {Y.}~\bibnamefont {Zhang}}, \bibinfo {author} {\bibfnamefont
  {P.~J.}\ \bibnamefont {Bustard}}, \ and\ \bibinfo {author} {\bibfnamefont
  {B.~J.}\ \bibnamefont {Sussman}},\ }\href@noop {} {\bibfield  {journal}
  {\bibinfo  {journal} {Applied Physics Letters}\ }\textbf {\bibinfo {volume}
  {119}},\ \bibinfo {pages} {160501} (\bibinfo {year} {2021})}\BibitemShut
  {NoStop}%
\bibitem [{\citenamefont {Donohue}\ \emph {et~al.}(2013)\citenamefont
  {Donohue}, \citenamefont {Agnew}, \citenamefont {Lavoie},\ and\ \citenamefont
  {Resch}}]{donohue2013coherent}%
  \BibitemOpen
  \bibfield  {author} {\bibinfo {author} {\bibfnamefont {J.~M.}\ \bibnamefont
  {Donohue}}, \bibinfo {author} {\bibfnamefont {M.}~\bibnamefont {Agnew}},
  \bibinfo {author} {\bibfnamefont {J.}~\bibnamefont {Lavoie}}, \ and\ \bibinfo
  {author} {\bibfnamefont {K.~J.}\ \bibnamefont {Resch}},\ }\href@noop {}
  {\bibfield  {journal} {\bibinfo  {journal} {Physical Review Letters}\
  }\textbf {\bibinfo {volume} {111}},\ \bibinfo {pages} {153602} (\bibinfo
  {year} {2013})}\BibitemShut {NoStop}%
\bibitem [{\citenamefont {Bouchard}\ \emph {et~al.}(2022)\citenamefont
  {Bouchard}, \citenamefont {England}, \citenamefont {Bustard}, \citenamefont
  {Heshami},\ and\ \citenamefont {Sussman}}]{bouchard2022quantum}%
  \BibitemOpen
  \bibfield  {author} {\bibinfo {author} {\bibfnamefont {F.}~\bibnamefont
  {Bouchard}}, \bibinfo {author} {\bibfnamefont {D.}~\bibnamefont {England}},
  \bibinfo {author} {\bibfnamefont {P.~J.}\ \bibnamefont {Bustard}}, \bibinfo
  {author} {\bibfnamefont {K.}~\bibnamefont {Heshami}}, \ and\ \bibinfo
  {author} {\bibfnamefont {B.}~\bibnamefont {Sussman}},\ }\href@noop {}
  {\bibfield  {journal} {\bibinfo  {journal} {PRX Quantum}\ }\textbf {\bibinfo
  {volume} {3}},\ \bibinfo {pages} {010332} (\bibinfo {year}
  {2022})}\BibitemShut {NoStop}%
\bibitem [{\citenamefont {Bouchard}\ \emph {et~al.}(2023)\citenamefont
  {Bouchard}, \citenamefont {Bonsma-Fisher}, \citenamefont {Heshami},
  \citenamefont {Bustard}, \citenamefont {England},\ and\ \citenamefont
  {Sussman}}]{bouchard2023measuring}%
  \BibitemOpen
  \bibfield  {author} {\bibinfo {author} {\bibfnamefont {F.}~\bibnamefont
  {Bouchard}}, \bibinfo {author} {\bibfnamefont {K.}~\bibnamefont
  {Bonsma-Fisher}}, \bibinfo {author} {\bibfnamefont {K.}~\bibnamefont
  {Heshami}}, \bibinfo {author} {\bibfnamefont {P.~J.}\ \bibnamefont
  {Bustard}}, \bibinfo {author} {\bibfnamefont {D.}~\bibnamefont {England}}, \
  and\ \bibinfo {author} {\bibfnamefont {B.}~\bibnamefont {Sussman}},\
  }\href@noop {} {\bibfield  {journal} {\bibinfo  {journal} {Physical Review
  A}\ }\textbf {\bibinfo {volume} {107}},\ \bibinfo {pages} {022618} (\bibinfo
  {year} {2023})}\BibitemShut {NoStop}%
\bibitem [{\citenamefont {Reck}\ \emph {et~al.}(1994)\citenamefont {Reck},
  \citenamefont {Zeilinger}, \citenamefont {Bernstein},\ and\ \citenamefont
  {Bertani}}]{reck1994experimental}%
  \BibitemOpen
  \bibfield  {author} {\bibinfo {author} {\bibfnamefont {M.}~\bibnamefont
  {Reck}}, \bibinfo {author} {\bibfnamefont {A.}~\bibnamefont {Zeilinger}},
  \bibinfo {author} {\bibfnamefont {H.~J.}\ \bibnamefont {Bernstein}}, \ and\
  \bibinfo {author} {\bibfnamefont {P.}~\bibnamefont {Bertani}},\ }\href@noop
  {} {\bibfield  {journal} {\bibinfo  {journal} {Physical review letters}\
  }\textbf {\bibinfo {volume} {73}},\ \bibinfo {pages} {58} (\bibinfo {year}
  {1994})}\BibitemShut {NoStop}%
\bibitem [{\citenamefont {Clements}\ \emph {et~al.}(2016)\citenamefont
  {Clements}, \citenamefont {Humphreys}, \citenamefont {Metcalf}, \citenamefont
  {Kolthammer},\ and\ \citenamefont {Walmsley}}]{clements2016optimal}%
  \BibitemOpen
  \bibfield  {author} {\bibinfo {author} {\bibfnamefont {W.~R.}\ \bibnamefont
  {Clements}}, \bibinfo {author} {\bibfnamefont {P.~C.}\ \bibnamefont
  {Humphreys}}, \bibinfo {author} {\bibfnamefont {B.~J.}\ \bibnamefont
  {Metcalf}}, \bibinfo {author} {\bibfnamefont {W.~S.}\ \bibnamefont
  {Kolthammer}}, \ and\ \bibinfo {author} {\bibfnamefont {I.~A.}\ \bibnamefont
  {Walmsley}},\ }\href@noop {} {\bibfield  {journal} {\bibinfo  {journal}
  {Optica}\ }\textbf {\bibinfo {volume} {3}},\ \bibinfo {pages} {1460}
  (\bibinfo {year} {2016})}\BibitemShut {NoStop}%
\bibitem [{\citenamefont {Spring}\ \emph {et~al.}(2013)\citenamefont {Spring},
  \citenamefont {Metcalf}, \citenamefont {Humphreys}, \citenamefont
  {Kolthammer}, \citenamefont {Jin}, \citenamefont {Barbieri}, \citenamefont
  {Datta}, \citenamefont {Thomas-Peter}, \citenamefont {Langford},
  \citenamefont {Kundys} \emph {et~al.}}]{spring2013boson}%
  \BibitemOpen
  \bibfield  {author} {\bibinfo {author} {\bibfnamefont {J.~B.}\ \bibnamefont
  {Spring}}, \bibinfo {author} {\bibfnamefont {B.~J.}\ \bibnamefont {Metcalf}},
  \bibinfo {author} {\bibfnamefont {P.~C.}\ \bibnamefont {Humphreys}}, \bibinfo
  {author} {\bibfnamefont {W.~S.}\ \bibnamefont {Kolthammer}}, \bibinfo
  {author} {\bibfnamefont {X.-M.}\ \bibnamefont {Jin}}, \bibinfo {author}
  {\bibfnamefont {M.}~\bibnamefont {Barbieri}}, \bibinfo {author}
  {\bibfnamefont {A.}~\bibnamefont {Datta}}, \bibinfo {author} {\bibfnamefont
  {N.}~\bibnamefont {Thomas-Peter}}, \bibinfo {author} {\bibfnamefont {N.~K.}\
  \bibnamefont {Langford}}, \bibinfo {author} {\bibfnamefont {D.}~\bibnamefont
  {Kundys}},  \emph {et~al.},\ }\href@noop {} {\bibfield  {journal} {\bibinfo
  {journal} {Science}\ }\textbf {\bibinfo {volume} {339}},\ \bibinfo {pages}
  {798} (\bibinfo {year} {2013})}\BibitemShut {NoStop}%
\bibitem [{\citenamefont {Ant{\'o}n}\ \emph {et~al.}(2019)\citenamefont
  {Ant{\'o}n}, \citenamefont {Loredo}, \citenamefont {Coppola}, \citenamefont
  {Ollivier}, \citenamefont {Viggianiello}, \citenamefont {Harouri},
  \citenamefont {Somaschi}, \citenamefont {Crespi}, \citenamefont {Sagnes},
  \citenamefont {Lemaitre} \emph {et~al.}}]{anton2019interfacing}%
  \BibitemOpen
  \bibfield  {author} {\bibinfo {author} {\bibfnamefont {C.}~\bibnamefont
  {Ant{\'o}n}}, \bibinfo {author} {\bibfnamefont {J.~C.}\ \bibnamefont
  {Loredo}}, \bibinfo {author} {\bibfnamefont {G.}~\bibnamefont {Coppola}},
  \bibinfo {author} {\bibfnamefont {H.}~\bibnamefont {Ollivier}}, \bibinfo
  {author} {\bibfnamefont {N.}~\bibnamefont {Viggianiello}}, \bibinfo {author}
  {\bibfnamefont {A.}~\bibnamefont {Harouri}}, \bibinfo {author} {\bibfnamefont
  {N.}~\bibnamefont {Somaschi}}, \bibinfo {author} {\bibfnamefont
  {A.}~\bibnamefont {Crespi}}, \bibinfo {author} {\bibfnamefont
  {I.}~\bibnamefont {Sagnes}}, \bibinfo {author} {\bibfnamefont
  {A.}~\bibnamefont {Lemaitre}},  \emph {et~al.},\ }\href@noop {} {\bibfield
  {journal} {\bibinfo  {journal} {Optica}\ }\textbf {\bibinfo {volume} {6}},\
  \bibinfo {pages} {1471} (\bibinfo {year} {2019})}\BibitemShut {NoStop}%
\bibitem [{\citenamefont {Wang}\ \emph {et~al.}(2019)\citenamefont {Wang},
  \citenamefont {Qin}, \citenamefont {Ding}, \citenamefont {Chen},
  \citenamefont {Chen}, \citenamefont {You}, \citenamefont {He}, \citenamefont
  {Jiang}, \citenamefont {You}, \citenamefont {Wang} \emph
  {et~al.}}]{wang2019boson}%
  \BibitemOpen
  \bibfield  {author} {\bibinfo {author} {\bibfnamefont {H.}~\bibnamefont
  {Wang}}, \bibinfo {author} {\bibfnamefont {J.}~\bibnamefont {Qin}}, \bibinfo
  {author} {\bibfnamefont {X.}~\bibnamefont {Ding}}, \bibinfo {author}
  {\bibfnamefont {M.-C.}\ \bibnamefont {Chen}}, \bibinfo {author}
  {\bibfnamefont {S.}~\bibnamefont {Chen}}, \bibinfo {author} {\bibfnamefont
  {X.}~\bibnamefont {You}}, \bibinfo {author} {\bibfnamefont {Y.-M.}\
  \bibnamefont {He}}, \bibinfo {author} {\bibfnamefont {X.}~\bibnamefont
  {Jiang}}, \bibinfo {author} {\bibfnamefont {L.}~\bibnamefont {You}}, \bibinfo
  {author} {\bibfnamefont {Z.}~\bibnamefont {Wang}},  \emph {et~al.},\
  }\href@noop {} {\bibfield  {journal} {\bibinfo  {journal} {Physical review
  letters}\ }\textbf {\bibinfo {volume} {123}},\ \bibinfo {pages} {250503}
  (\bibinfo {year} {2019})}\BibitemShut {NoStop}%
\bibitem [{\citenamefont {Zhong}\ \emph {et~al.}(2021)\citenamefont {Zhong},
  \citenamefont {Deng}, \citenamefont {Qin}, \citenamefont {Wang},
  \citenamefont {Chen}, \citenamefont {Peng}, \citenamefont {Luo},
  \citenamefont {Wu}, \citenamefont {Gong}, \citenamefont {Su} \emph
  {et~al.}}]{zhong2021phase}%
  \BibitemOpen
  \bibfield  {author} {\bibinfo {author} {\bibfnamefont {H.-S.}\ \bibnamefont
  {Zhong}}, \bibinfo {author} {\bibfnamefont {Y.-H.}\ \bibnamefont {Deng}},
  \bibinfo {author} {\bibfnamefont {J.}~\bibnamefont {Qin}}, \bibinfo {author}
  {\bibfnamefont {H.}~\bibnamefont {Wang}}, \bibinfo {author} {\bibfnamefont
  {M.-C.}\ \bibnamefont {Chen}}, \bibinfo {author} {\bibfnamefont {L.-C.}\
  \bibnamefont {Peng}}, \bibinfo {author} {\bibfnamefont {Y.-H.}\ \bibnamefont
  {Luo}}, \bibinfo {author} {\bibfnamefont {D.}~\bibnamefont {Wu}}, \bibinfo
  {author} {\bibfnamefont {S.-Q.}\ \bibnamefont {Gong}}, \bibinfo {author}
  {\bibfnamefont {H.}~\bibnamefont {Su}},  \emph {et~al.},\ }\href@noop {}
  {\bibfield  {journal} {\bibinfo  {journal} {Physical review letters}\
  }\textbf {\bibinfo {volume} {127}},\ \bibinfo {pages} {180502} (\bibinfo
  {year} {2021})}\BibitemShut {NoStop}%
\bibitem [{\citenamefont {Kupchak}\ \emph {et~al.}(2019)\citenamefont
  {Kupchak}, \citenamefont {Erskine}, \citenamefont {England},\ and\
  \citenamefont {Sussman}}]{kupchak2019terahertz}%
  \BibitemOpen
  \bibfield  {author} {\bibinfo {author} {\bibfnamefont {C.}~\bibnamefont
  {Kupchak}}, \bibinfo {author} {\bibfnamefont {J.}~\bibnamefont {Erskine}},
  \bibinfo {author} {\bibfnamefont {D.}~\bibnamefont {England}}, \ and\
  \bibinfo {author} {\bibfnamefont {B.}~\bibnamefont {Sussman}},\ }\href@noop
  {} {\bibfield  {journal} {\bibinfo  {journal} {Optics letters}\ }\textbf
  {\bibinfo {volume} {44}},\ \bibinfo {pages} {1427} (\bibinfo {year}
  {2019})}\BibitemShut {NoStop}%
\bibitem [{\citenamefont {Brandt}\ \emph {et~al.}(2020)\citenamefont {Brandt},
  \citenamefont {Hiekkam{\"a}ki}, \citenamefont {Bouchard}, \citenamefont
  {Huber},\ and\ \citenamefont {Fickler}}]{brandt2020high}%
  \BibitemOpen
  \bibfield  {author} {\bibinfo {author} {\bibfnamefont {F.}~\bibnamefont
  {Brandt}}, \bibinfo {author} {\bibfnamefont {M.}~\bibnamefont
  {Hiekkam{\"a}ki}}, \bibinfo {author} {\bibfnamefont {F.}~\bibnamefont
  {Bouchard}}, \bibinfo {author} {\bibfnamefont {M.}~\bibnamefont {Huber}}, \
  and\ \bibinfo {author} {\bibfnamefont {R.}~\bibnamefont {Fickler}},\
  }\href@noop {} {\bibfield  {journal} {\bibinfo  {journal} {Optica}\ }\textbf
  {\bibinfo {volume} {7}},\ \bibinfo {pages} {98} (\bibinfo {year}
  {2020})}\BibitemShut {NoStop}%
\bibitem [{\citenamefont {Chi}\ \emph {et~al.}(2022)\citenamefont {Chi},
  \citenamefont {Huang}, \citenamefont {Zhang}, \citenamefont {Mao},
  \citenamefont {Zhou}, \citenamefont {Chen}, \citenamefont {Zhai},
  \citenamefont {Bao}, \citenamefont {Dai}, \citenamefont {Yuan} \emph
  {et~al.}}]{chi2022programmable}%
  \BibitemOpen
  \bibfield  {author} {\bibinfo {author} {\bibfnamefont {Y.}~\bibnamefont
  {Chi}}, \bibinfo {author} {\bibfnamefont {J.}~\bibnamefont {Huang}}, \bibinfo
  {author} {\bibfnamefont {Z.}~\bibnamefont {Zhang}}, \bibinfo {author}
  {\bibfnamefont {J.}~\bibnamefont {Mao}}, \bibinfo {author} {\bibfnamefont
  {Z.}~\bibnamefont {Zhou}}, \bibinfo {author} {\bibfnamefont {X.}~\bibnamefont
  {Chen}}, \bibinfo {author} {\bibfnamefont {C.}~\bibnamefont {Zhai}}, \bibinfo
  {author} {\bibfnamefont {J.}~\bibnamefont {Bao}}, \bibinfo {author}
  {\bibfnamefont {T.}~\bibnamefont {Dai}}, \bibinfo {author} {\bibfnamefont
  {H.}~\bibnamefont {Yuan}},  \emph {et~al.},\ }\href@noop {} {\bibfield
  {journal} {\bibinfo  {journal} {Nature communications}\ }\textbf {\bibinfo
  {volume} {13}},\ \bibinfo {pages} {1166} (\bibinfo {year}
  {2022})}\BibitemShut {NoStop}%
\bibitem [{\citenamefont {Ringbauer}\ \emph {et~al.}(2022)\citenamefont
  {Ringbauer}, \citenamefont {Meth}, \citenamefont {Postler}, \citenamefont
  {Stricker}, \citenamefont {Blatt}, \citenamefont {Schindler},\ and\
  \citenamefont {Monz}}]{ringbauer2022universal}%
  \BibitemOpen
  \bibfield  {author} {\bibinfo {author} {\bibfnamefont {M.}~\bibnamefont
  {Ringbauer}}, \bibinfo {author} {\bibfnamefont {M.}~\bibnamefont {Meth}},
  \bibinfo {author} {\bibfnamefont {L.}~\bibnamefont {Postler}}, \bibinfo
  {author} {\bibfnamefont {R.}~\bibnamefont {Stricker}}, \bibinfo {author}
  {\bibfnamefont {R.}~\bibnamefont {Blatt}}, \bibinfo {author} {\bibfnamefont
  {P.}~\bibnamefont {Schindler}}, \ and\ \bibinfo {author} {\bibfnamefont
  {T.}~\bibnamefont {Monz}},\ }\href@noop {} {\bibfield  {journal} {\bibinfo
  {journal} {Nature Physics}\ }\textbf {\bibinfo {volume} {18}},\ \bibinfo
  {pages} {1053} (\bibinfo {year} {2022})}\BibitemShut {NoStop}%
\bibitem [{\citenamefont {Matsuda}(2016)}]{matsuda2016deterministic}%
  \BibitemOpen
  \bibfield  {author} {\bibinfo {author} {\bibfnamefont {N.}~\bibnamefont
  {Matsuda}},\ }\href@noop {} {\bibfield  {journal} {\bibinfo  {journal}
  {Science advances}\ }\textbf {\bibinfo {volume} {2}},\ \bibinfo {pages}
  {e1501223} (\bibinfo {year} {2016})}\BibitemShut {NoStop}%
\bibitem [{\citenamefont {McGuinness}\ \emph {et~al.}(2010)\citenamefont
  {McGuinness}, \citenamefont {Raymer}, \citenamefont {McKinstrie},\ and\
  \citenamefont {Radic}}]{mcguinness2010quantum}%
  \BibitemOpen
  \bibfield  {author} {\bibinfo {author} {\bibfnamefont {H.~J.}\ \bibnamefont
  {McGuinness}}, \bibinfo {author} {\bibfnamefont {M.~G.}\ \bibnamefont
  {Raymer}}, \bibinfo {author} {\bibfnamefont {C.~J.}\ \bibnamefont
  {McKinstrie}}, \ and\ \bibinfo {author} {\bibfnamefont {S.}~\bibnamefont
  {Radic}},\ }\href@noop {} {\bibfield  {journal} {\bibinfo  {journal}
  {Physical review letters}\ }\textbf {\bibinfo {volume} {105}},\ \bibinfo
  {pages} {093604} (\bibinfo {year} {2010})}\BibitemShut {NoStop}%
\bibitem [{\citenamefont {Joshi}\ \emph {et~al.}(2018)\citenamefont {Joshi},
  \citenamefont {Farsi}, \citenamefont {Clemmen}, \citenamefont {Ramelow},\
  and\ \citenamefont {Gaeta}}]{joshi2018frequency}%
  \BibitemOpen
  \bibfield  {author} {\bibinfo {author} {\bibfnamefont {C.}~\bibnamefont
  {Joshi}}, \bibinfo {author} {\bibfnamefont {A.}~\bibnamefont {Farsi}},
  \bibinfo {author} {\bibfnamefont {S.}~\bibnamefont {Clemmen}}, \bibinfo
  {author} {\bibfnamefont {S.}~\bibnamefont {Ramelow}}, \ and\ \bibinfo
  {author} {\bibfnamefont {A.~L.}\ \bibnamefont {Gaeta}},\ }\href@noop {}
  {\bibfield  {journal} {\bibinfo  {journal} {Nature communications}\ }\textbf
  {\bibinfo {volume} {9}},\ \bibinfo {pages} {847} (\bibinfo {year}
  {2018})}\BibitemShut {NoStop}%
\bibitem [{\citenamefont {Korzh}\ \emph {et~al.}(2020)\citenamefont {Korzh},
  \citenamefont {Zhao}, \citenamefont {Allmaras}, \citenamefont {Frasca},
  \citenamefont {Autry}, \citenamefont {Bersin}, \citenamefont {Beyer},
  \citenamefont {Briggs}, \citenamefont {Bumble}, \citenamefont {Colangelo}
  \emph {et~al.}}]{korzh2020demonstration}%
  \BibitemOpen
  \bibfield  {author} {\bibinfo {author} {\bibfnamefont {B.}~\bibnamefont
  {Korzh}}, \bibinfo {author} {\bibfnamefont {Q.-Y.}\ \bibnamefont {Zhao}},
  \bibinfo {author} {\bibfnamefont {J.~P.}\ \bibnamefont {Allmaras}}, \bibinfo
  {author} {\bibfnamefont {S.}~\bibnamefont {Frasca}}, \bibinfo {author}
  {\bibfnamefont {T.~M.}\ \bibnamefont {Autry}}, \bibinfo {author}
  {\bibfnamefont {E.~A.}\ \bibnamefont {Bersin}}, \bibinfo {author}
  {\bibfnamefont {A.~D.}\ \bibnamefont {Beyer}}, \bibinfo {author}
  {\bibfnamefont {R.~M.}\ \bibnamefont {Briggs}}, \bibinfo {author}
  {\bibfnamefont {B.}~\bibnamefont {Bumble}}, \bibinfo {author} {\bibfnamefont
  {M.}~\bibnamefont {Colangelo}},  \emph {et~al.},\ }\href@noop {} {\bibfield
  {journal} {\bibinfo  {journal} {Nature Photonics}\ }\textbf {\bibinfo
  {volume} {14}},\ \bibinfo {pages} {250} (\bibinfo {year} {2020})}\BibitemShut
  {NoStop}%
\bibitem [{\citenamefont {Steinbrecher}\ \emph {et~al.}(2019)\citenamefont
  {Steinbrecher}, \citenamefont {Olson}, \citenamefont {Englund},\ and\
  \citenamefont {Carolan}}]{steinbrecher2019quantum}%
  \BibitemOpen
  \bibfield  {author} {\bibinfo {author} {\bibfnamefont {G.~R.}\ \bibnamefont
  {Steinbrecher}}, \bibinfo {author} {\bibfnamefont {J.~P.}\ \bibnamefont
  {Olson}}, \bibinfo {author} {\bibfnamefont {D.}~\bibnamefont {Englund}}, \
  and\ \bibinfo {author} {\bibfnamefont {J.}~\bibnamefont {Carolan}},\
  }\href@noop {} {\bibfield  {journal} {\bibinfo  {journal} {npj Quantum
  Information}\ }\textbf {\bibinfo {volume} {5}},\ \bibinfo {pages} {60}
  (\bibinfo {year} {2019})}\BibitemShut {NoStop}%
\bibitem [{\citenamefont {Matthews}\ \emph {et~al.}(2016)\citenamefont
  {Matthews}, \citenamefont {Zhou}, \citenamefont {Cable}, \citenamefont
  {Shadbolt}, \citenamefont {Saunders}, \citenamefont {Durkin}, \citenamefont
  {Pryde},\ and\ \citenamefont {O’Brien}}]{matthews2016towards}%
  \BibitemOpen
  \bibfield  {author} {\bibinfo {author} {\bibfnamefont {J.~C.}\ \bibnamefont
  {Matthews}}, \bibinfo {author} {\bibfnamefont {X.-Q.}\ \bibnamefont {Zhou}},
  \bibinfo {author} {\bibfnamefont {H.}~\bibnamefont {Cable}}, \bibinfo
  {author} {\bibfnamefont {P.~J.}\ \bibnamefont {Shadbolt}}, \bibinfo {author}
  {\bibfnamefont {D.~J.}\ \bibnamefont {Saunders}}, \bibinfo {author}
  {\bibfnamefont {G.~A.}\ \bibnamefont {Durkin}}, \bibinfo {author}
  {\bibfnamefont {G.~J.}\ \bibnamefont {Pryde}}, \ and\ \bibinfo {author}
  {\bibfnamefont {J.~L.}\ \bibnamefont {O’Brien}},\ }\href@noop {} {\bibfield
   {journal} {\bibinfo  {journal} {npj Quantum Information}\ }\textbf {\bibinfo
  {volume} {2}},\ \bibinfo {pages} {1} (\bibinfo {year} {2016})}\BibitemShut
  {NoStop}%
\bibitem [{\citenamefont {Cacciapuoti}\ \emph {et~al.}(2019)\citenamefont
  {Cacciapuoti}, \citenamefont {Caleffi}, \citenamefont {Tafuri}, \citenamefont
  {Cataliotti}, \citenamefont {Gherardini},\ and\ \citenamefont
  {Bianchi}}]{cacciapuoti2019quantum}%
  \BibitemOpen
  \bibfield  {author} {\bibinfo {author} {\bibfnamefont {A.~S.}\ \bibnamefont
  {Cacciapuoti}}, \bibinfo {author} {\bibfnamefont {M.}~\bibnamefont
  {Caleffi}}, \bibinfo {author} {\bibfnamefont {F.}~\bibnamefont {Tafuri}},
  \bibinfo {author} {\bibfnamefont {F.~S.}\ \bibnamefont {Cataliotti}},
  \bibinfo {author} {\bibfnamefont {S.}~\bibnamefont {Gherardini}}, \ and\
  \bibinfo {author} {\bibfnamefont {G.}~\bibnamefont {Bianchi}},\ }\href@noop
  {} {\bibfield  {journal} {\bibinfo  {journal} {IEEE Network}\ }\textbf
  {\bibinfo {volume} {34}},\ \bibinfo {pages} {137} (\bibinfo {year}
  {2019})}\BibitemShut {NoStop}%
\end{thebibliography}
\end{document}